\documentclass[submission,copyright,creativecommons]{eptcs}
\providecommand{\event}{HCVS 2022}
\usepackage{breakurl}             
\usepackage{underscore}  
\usepackage[utf8]{inputenc}
\usepackage[T1]{fontenc}
\usepackage[english]{babel}
\usepackage{amsthm}
\usepackage{amsmath}
\usepackage{amsfonts}
\usepackage{amssymb}
\usepackage{array}
\usepackage{breakcites}
\usepackage{xspace}
\usepackage{enumitem}
\usepackage{color}
\usepackage{wrapfig}  
\usepackage{colortbl} 
\usepackage{subcaption}   
\usepackage{url}
\usepackage{stmaryrd}
\usepackage{hyperref}
\DeclareMathAlphabet{\mathcal}{OMS}{cmsy}{m}{n}

\usepackage{listings}

\lstdefinestyle{myscala}{
	language=scala,
	stringstyle=\ttfamily,
	showstringspaces = false,
	basicstyle=\linespread{0.9}\small\ttfamily,
	commentstyle=\small\emph,
	keywordstyle=\color{blue}\bfseries,
	mathescape=true,
	breaklines=true,
	xleftmargin=0em, 
	,morekeywords={ensuring,require}
	,otherkeywords={}
	,deletekeywords={true,false},
	columns=[l]flexible
}

\lstdefinestyle{myprolog}{
	,style=myscala,
	,language=prolog
	,deletekeywords={false,true}
}

\title{Contract Strengthening \\through Constrained Horn Clause Verification}

\author{
	Emanuele De Angelis
	\institute{IASI-CNR, Italy}
	\email{emanuele.deangelis@iasi.cnr.it}
	\and
	Fabio Fioravanti
	\institute{DEc, University of Chieti-Pescara, Italy}
	\email{fabio.fioravanti@unich.it}
	\and
	Alberto Pettorossi
	\institute{DICII, University of Rome `Tor Vergata', Italy}
	\email{pettorossi@info.uniroma2.it}
	\and
	Maurizio Proietti
	\institute{IASI-CNR, Italy}
	\email{maurizio.proietti@iasi.cnr.it}
}

\def\titlerunning{Contract Strengthening through Constrained Horn Clause Verification}

\def\authorrunning{E.~De Angelis, F.~Fioravanti, A.~Pettorossi, and M.~Proietti}

\renewcommand{\em}{\it}   

\usepackage{proof}

\newcommand{\tts}[1]{{\small{\texttt{#1}}}}
\newcommand{\ttsi}[1]{{\small{\textsf{\textit{{#1}}}}}}

\newcommand{\Cata}{${\mathcal T}_{\mathit{cata}}$}

\newtheorem{proof*}{Proof}   

\newcommand{\vericat}{\mbox{{VeriCa\hspace{-0.3mm}T}}\xspace}

\newenvironment{sizepar}[2]
{\par\fontsize{#1}{#2}\selectfont}
{\par}

\begin{document}
\maketitle

\newcommand{\TCata}{$\mathit{T\!_{cata}}$}   

\label{sec:abstract}
\begin{abstract}
The functional properties of a program are often specified by providing a contract for each of its functions.
A contract of a function consists of a pair of formulas, called a precondition and a postcondition, which, respectively, should hold before and after execution of that function.
It might be the case that the contracts supplied by the programmer are not adequate to allow a verification system to prove program correctness, that is, to show that for every function, if the precondition holds and the execution of the function terminates, then the postcondition holds.
We address this problem by providing a technique which may strengthen the postconditions of the functions, thereby improving the ability of the verifier to show program correctness.
Our technique consists of four steps.
First, the translation of the given program, which may manipulate algebraic data structures (ADTs), and its contracts into a set of constrained Horn clauses (CHCs) whose satisfiability implies the validity of the given contracts.
Then, the derivation, via CHC transformation performed by the \vericat tool, of a new set of CHCs that manipulate only basic sorts (such as booleans or integers) and whose satisfiability implies the satisfiability of the original set of clauses.
Then, the construction of a model, if any, of the new, derived CHCs using the CHC solver SPACER for basic sorts.
Finally, the translation of that model into the formulas that suitably strengthen the postconditions of the given contracts.
We will present our technique through an example consisting of a Scala program for reversing lists.
Note that the {\sc Stainless} verifier is not able to prove the correctness of that program when considering the given contracts, while it succeeds when considering the contracts with the strengthened postconditions constructed by applying our technique.
\end{abstract}

\section{Introduction}
\label{sec:Intro}

In many program verification techniques \`{a} la Floyd-Hoare~\cite{Flo67,Hoa69},
the meaning of a program is specified by providing a {\em contract},
that is, a pair of 
 a {\it precondition} and a {\it postcondition} formula for each of the program functions.
A program function is said to be {\it partially correct} with respect to a given
contract if~the precondition holds before function execution, and 
the function terminates, then the postcondition holds. A program function is said to be {\it totally correct} if it is partially correct and it terminates whenever its precondition holds.
Many programming languages 
(for instance, Ada~\cite{BoochB94}, Ciao~\cite{Hermenegildo&12}, 
Eiffel~\cite{Meyer91}, Scala~\cite{OderskySV11}, and Solidity~\cite{Solidity})
provide support for contract specification.

Programmers write contracts to specify invariant properties of entire programs
or program fragments (such as functions, methods, and loops)
and these contracts 
may be used by verifiers (e.g., {\sc Boogie}~\cite{Boogie}, {\sc Leon}~\cite{Su&11}, {\sc Why3}~\cite{Why3}, {\sc Dafny}~\cite{Lei13},  and {\sc Stainless}~\cite{HamzaVK19})
to generate and possibly prove suitable verification conditions, that is, formulas
whose validity guarantees program correctness. 

Verification conditions are usually proved by using theorem provers  
or Satisfiability Modulo Theory (SMT) 
solvers\,\cite{CVC4,MaS13,HoR18,DeB08} and 
{constrained Horn clause} (CHC) solvers, such as Eldarica\,\cite{HoR18} 
and {\sc{SPACER}}\,\cite{Ko&14}.\ In these solvers there is support 
for a wide range of logical theories, 
including basic data types, such as integers and booleans,
and also more complex data structures.

In the case of programs manipulating complex data
structures, such as Algebraic Data Types (ADTs), 
loop invariants and contracts may be quite complicated
and their verification may require the enhancement of the solvers 
by incorporating inductive proof rules~\cite{ReK15,Un&17,Ya&19}, or
tree automata-based techniques~\cite{KostyukovMF21}, or 
CHC abstractions~\cite{GovindSG22}.

Besides the presence of ADTs, there is often one additional reason  
that makes it difficult to prove the validity of the contracts
when using an automated verification system. It is the fact that 
the contracts specified by the programmer are not sufficiently detailed
in specifying the behaviour of the functions. 

In order to clarify this point, let us consider the case of
a program made out of some functions which may call each other,
possibly in a mutually recursive way. Each of these functions has its
own contract. Now it may happen that a program verifier
is able to prove the contract of a particular function, say $g$, 
while it fails 
to prove the contract of another function, say $f$, because $f$ calls
 $g$ and the 
contract that has been proved for $g$ is not detailed enough.
This may happen because the programmer, when specifying contracts,
did not take into account the fact 
that~$g$ is called by $f$\!, and for the proof of the contract of~$f$ 
a stronger, more detailed contract for $g$ is indeed required.
Moreover, these contract 
interdependencies are not always easy to take into 
account by the programmer, and this happens in particular when
 the program is made out of many function definitions.

In this paper we present a novel technique for 
strengthening the function contracts in the case of programs that manipulate ADTs, 
and then, our technique allows an easier proof of contract correctness. Our technique consists of four steps.


{\rm{Step~(i)}}:~First,~we translate the given program with its contracts into a set
of CHCs, so that the contract verification problem is translated 
into an equivalent satisfiability problem for that set of CHCs.

{\rm{Step~(ii)}}:~Then, by using already known methods~\cite{De&18a,De&20a}, we transform
this set of CHCs into a new set where all ADT terms are removed. In  particular, we use
the \vericat tool~\cite{DeAngelisFPP22}.
These methods are sound, in the sense that
the satisfiability of the transformed clauses implies the satisfiability of the original set of clauses. Under suitable hypotheses, these methods are also complete, that is, 
the original and the transformed set of clauses are equisatisfiable~\cite{De&22a}.
In this way, we separate the concern of dealing
with ADTs (at transformation time) from the concern of dealing with
simpler, non-inductive constraint theories (at solving time), thus
avoiding the complex interaction between inductive reasoning and 
constraint solving. Usually, during this transformation new predicate symbols are introduced in the derived CHCs. 

{\rm{Step~(iii)}}:~Then, we invoke a
CHC solver (in our case {\sc{SPACER}}) to show the satisfiability of the new, transformed set of CHCs and to construct the models, if any, of the new predicates which have been introduced.  

{\rm{Step~(iv)}}:~Finally, from those models 
we construct the formulas which strengthen the given contracts.

\smallskip
The viability and the power of this novel technique will be shown in the following sections
through a simple example dealing with lists.

\section{Verification of Program Contracts}
\label{sec:VerifContracts}

In this section we present the program verification problem we consider with the help of an example.
Let us consider the Scala program {\ttsi{Reverse}},  depicted in Figure~\ref{fig:RevScala},
for computing the reversal of a list.
In that program the 
function preconditions and postconditions are specified by \tts{require}
and \tts{ensuring} assertions, respectively.
%
The contract for the function \tts{rev} states that,
if a list~\tts{l} of integers is sorted in {\it ascending} order (w.r.t.~$\leq$),
then the list \tts{rev(l)} is sorted in {\it descending} order (that is, it is sorted w.r.t.~$\geq$).
The ascending (or descending) order 
 for list~{\tts{l}} 
is checked by the function 
\tts{is\_asorted(l)} (or \mbox{\tts{is\_dsorted(l)}}, respectively). 
The contract for the function \tts{snoc(l,x)},
which appends element~{\tts x} to the end of list {\tts l}, 
states that 
if list \tts{l} is sorted in descending order and 
\tts{leq\_all(x,l)} holds (that is, element {\tts{x}} is less than or equal 
to every element of {\tts l}),
then  also \tts{snoc(l,x)} is sorted in {descending} order. In program \ttsi{Reverse}
we also need the function \tts{hd} which, given a list of integers, returns a pair consisting of a
boolean and an integer. If the given list
is not empty and \tts {h} is its head, then the returned pair is 
\mbox{$\langle$\tts{true}, \tts{h}$\rangle$,} while it is $\langle$\tts{false}, \tts{0}$\rangle$, 
if the list is empty, being \tts{0} an arbitrary integer value (which is never used elsewhere in the program).

\begin{figure}[ht!]
\begin{sizepar}{9}{9.5}
\begin{verbatim}
object Reverse {
  def rev(l: List[BigInt]): List[BigInt] = {
    require(is_asorted(l))                              // precondition of rev 
    l match {
      case Nil() => Nil[BigInt]()
      case Cons(x, xs) => snoc(rev(xs),x) }
      } ensuring { res => is_dsorted(res) }             // postcondition of rev
      
  def snoc(l: List[BigInt], x:BigInt): List[BigInt] = {
    require(is_dsorted(l) && leq_all(x,l))              // precondition of snoc
    l match {
      case Nil() => Cons(x,Nil())
      case Cons(y, ys) => Cons(y,snoc(ys,x)) }
         } ensuring { res => is_dsorted(res) }          // postcondition of snoc
         
  def is_asorted(l: List[BigInt]): Boolean = {  
    l match {
      case Nil() => true
      case Cons(x,xs) => !(hd(xs)._1) || (x <= (hd(xs)._2) && is_asorted(xs))  } }
      
  def is_dsorted(l: List[BigInt]): Boolean = {   
      l match {
      case Nil() => true
      case Cons(x,xs) => !(hd(xs)._1) || (x >= (hd(xs)._2) && is_dsorted(xs))  } }
      
  def hd(l: List[BigInt]): (Boolean, BigInt) = {
    l match {
      case Nil() => (false, BigInt(0))
      case Cons(x, xs) => (true, x) } }
      
  def leq_all(x: BigInt, l: List[BigInt]): Boolean = {     
    l match {
      case Nil() => true
      case Cons(y, ys) => if (x > y) {false} else { leq_all(x, ys) } } }
}  // end of object Reverse
\end{verbatim}
\end{sizepar}
\caption{Program {\ttsi{Reverse}} with the
contracts for the functions \tts{rev} and \tts{snoc}. `\tts{!}', `\tts{\&\&}', and `\tts{||}' denote
boolean negation, conjunction, and disjunction, respectively. `\tts{p.\_1}', and `\tts{p.\_2}' denote
the first and the second projection of a given pair \tts{p}, respectively.
\label{fig:RevScala}}
\end{figure}

\noindent
In order to prove the validity of a contract 
$\langle$\tts{precond(x), postcond(x,f(x))}$\rangle$ for a given 
function \tts{f}, 
we need to prove that
\tts{$\forall$x.\;precond(x) $\rightarrow$ postcond(x,f(x))}.
Now, if we submit to {\sc Stainless}~\cite{HamzaVK19}, which is a verifier for Scala programs, the 
above \ttsi{Reverse} program with the initial directives:
\begin{sizepar}{10}{11}
\begin{verbatim}
    import stainless.proof._
    import stainless.lang._
    import stainless.collection._
\end{verbatim}
\end{sizepar}

\noindent
we get that {\sc Stainless} is not able to 
check the validity of the contract for \tts{rev}, because it fails to establish (within the
timeout of 100 s) the precondition for the function call `\tts{snoc(rev(xs),x)}'
(occurring in the \tts{Cons} case for \tts{rev}), which is needed to use the postcondition of \tts{snoc} and
prove that \tts{rev(Cons(x, xs))} is
sorted in descending order.  Indeed, {\sc Stainless} returns the following warning:\\
\hspace*{8mm}\tts{=> TIMEOUT case Cons(x, xs) => snoc(rev(xs),x)}

\noindent
In the next section we will see in action our technique for strengthening
contracts in the case of the \ttsi{Reverse} example. By using the strengthened contract for the function \tts{rev}, {\sc Stainless} is indeed able
to construct, as desired, a proof of the validity of both contracts of that
example, that is, the validity of the contracts for the
functions \tts{rev} and \tts{snoc}.


\section{Verifying Contracts via CHC Satisfiability}
\label{sec:VerifContractsCHCSatisf}
As already mentioned at the end of Section~\ref{sec:Intro} 
our technique for strengthening the contracts is made out of four steps
which we now perform in the case of the \ttsi{Reverse} program shown in
Figure~\ref{fig:RevScala}.

Step~(i).  We translate the given Scala program and its contracts into the set \ttsi{ReverseCHCs} of clauses shown 
in Figure~\ref{fig:RevCHCs}. In this translation we maintain the {\it call-by-value} semantics
of the Scala program in the sense that, if a function \tts{f} applied to the input \tts{X}
evaluates to output \tts{Y}, 
then the atom \tts{f(X,Y)} 
occurs in the least model of the
set \ttsi{ReverseCHCs}.

Several techniques for translating imperative and functional programs have been defined 
in the literature~\cite{DeAngelisFGHPP21,Gr&12}.
Even if no specific tool is available, 
we may assume that, by applying one of those techniques
we get a translation of Scala functions to CHC predicates 
that guarantees that the set \mbox{\ttsi{ReverseCHCs}}  
is satisfiable if and only if
the contracts for the functions \tts{rev} and \tts{snoc} of the program 
\ttsi{Reverse} are valid.
%
%
%
In Figure~\ref{fig:RevCHCs}, the contracts for the functions \tts{rev} and \tts{snoc} are encoded 
by the two constrained goals~{\tts{GR}} and~{\tts{GS}},
respectively. Here and in what follows, we call `constrained goal' (or `goal', for short) 
any clause whose head is \tts{false}. 

Note that when writing clauses, we often prefer writing 
the variable \tts{X}, 
instead of the constraint \tts{X=true}, and the negated variable \tts{\textasciitilde X}, instead of the constraint \tts{X=false}.
In particular, in clause \tts{GR} of Figure~\ref{fig:RevCHCs} we have written \tts{(BL} \tts{\,\&\,} 
\tts{\textasciitilde BR)}, instead of the equivalent constraint \tts{(BL=true} \tts{\,\&\,} \tts{BR=false)}.

\begin{figure}[ht!]
\begin{sizepar}{9}{10}
%
\begin{verbatim}
/* --------- CHC translation of the functions rev and snoc --------- */
    rev([],[]).                                  
    rev([H|T],R) :- rev(T,S), snoc(S,H,R).       
    snoc([],X,[X]).
    snoc([X|Xs],Y,[X|Zs]) :- snoc(Xs,Y,Zs).
    
/* --------- CHC translation of the functions used by the contracts  --------- */
    is_asorted([],Res) :- Res. 
    is_asorted([X|Xs],Res) :- Res = (IsDefXs => (X=<HdXs & ResXs)), 
                              hd(Xs,IsDefXs,HdXs), is_asorted(Xs,ResXs). 
    is_dsorted([],Res) :- Res. 
    is_dsorted([X|Xs],Res) :- Res = (IsDefXs => (X>=HdXs & ResXs)), 
                              hd(Xs,IsDefXs,HdXs), is_dsorted(Xs,ResXs). 
    hd([],IsDef,Hd) :- ~IsDef & Hd=0.            
    hd([H|T],IsDef,Hd) :- IsDef & Hd=H.  
    leq_all(N,[],B) :- B.                   
    leq_all(N,[X|Xs],B) :- B = (N=<X & B1), leq_all(N,Xs,B1).   
    
/* --------- CHC translation of the contracts of the functions rev and snoc --------- */
GR. false :- (BL & ~BR), rev(L,R), is_asorted(L,BL), is_dsorted(R,BR).
GS. false :- (BX & BA & ~BC), snoc(A,X,C), is_dsorted(A,BA), 
              leq_all(X,A,BX), is_dsorted(C,BC).
\end{verbatim} 
\end{sizepar}
\vspace{-2mm}
\caption{The set {\ttsi{ReverseCHCs}} of clauses 
for the  program {\ttsi{Reverse}}. 
In constraint formulas we use integer and boolean variables, the predicate 
`\tts{=}'~(equality) and the operators `\tts{\textasciitilde}' (negation),
`\tts{\&}'~(conjunction), and `\tts{=>}' (implication). \label{fig:RevCHCs}}
	\vspace{-2mm}
\end{figure}

Now, in order to show the validity of the contracts for \tts{rev} and \tts{snoc},
we have to show that the set \ttsi{ReverseCHCs} of clauses
is satisfiable.
Unfortunately, the CHC solvers Eldarica and SPACER
are not capable to solve this satisfiability problem. 
This is basically due to the fact that those solvers
lack any form of inductive reasoning on lists and, moreover, they
do not use the information about the validity of the contract for \tts{snoc}  during the proof
of satisfiability of the goal which encodes the contract for \tts{rev}.

Then, we proceed according to Step~(ii) of our technique, which consists in 
applying a transformation that removes all ADT terms from \ttsi{ReverseCHCs}.
Indeed, we apply Algorithm~\Cata~\cite{DeAngelisFPP22},
implemented in the~\vericat tool, and we get
the new set {\ttsi{TransfReverseCHCs}} of clauses (see Figure~\ref{fig:TransfReverseCHCs-noADTs}), whose satisfibility implies the
satisfiability of the set {\ttsi{ReverseCHCs}}.
In particular, starting from goal~\tts{GR}, by transformation we obtain clauses \tts{T1}--\tts{T5}, and 
starting goal~\tts{GS}, by transformation we obtain clauses \tts{T6}--\tts{T8}.

Let us make a minor remark about the numbering of 
the new predicates occurring in clauses \tts{T1}--\tts{T5} 
obtained by clause transformations starting from goal~\tts{GR}. 
(A similar remark applies to clauses \mbox{\tts{T6}--\tts{T8}} when starting from goal~\tts{GS}.)
\vericat uses for the new predicates to be introduced,
a progressive numbering starting from the name \tts{new1}, that is, it uses \tts{new1}, \tts{new2}, and so 
on.  
Now only predicates \tts{new3} and \tts{new7}
occur in clauses \tts{T1}--\tts{T5} because of the following two 
reasons: 
(i)~during transformation, some of the new predicates that are introduced,
are generalizations of already introduced ones, and thus one can replace the 
less general predicates by the more general ones, 
and
(ii)~some atoms with new predicates are unfolded during clause transformation, and thus they will not occur 
in the derived final clauses.

\begin{figure}[ht!]
\vspace{1mm}
\begin{sizepar}{9}{10}
\begin{verbatim}
/* --------- T1-T5: clauses derived for the rev contract GR  --------- */
T1. new7(A,B,C,D,E,F,G,H,D,I,J) :- A & B=D & C=(K=>((D>=L)&M)) & E & ~F & G=0 & H &
                     & J=((I=<D)&N) & M & ~K & L=0 & N.
T2. new7(A,B,C,D,E,F,G,H,D,I,J) :- A & B=K & C=(L=>((K>=M)&N)) & E=((D=<K)&T) & F & G=K & 
                     & H=(P=>((K>=Q)&R)) & J=((I=<K)&S) & (R&T)=>N, new7(L,M,N,D,T,P,Q,R,D,I,S).
T3. new3(A,B,C,D,E,F) :- A & C & ~D & E=0 & F.
T4. new3(A,B,C,D,E,F) :- D & E=G & F=(H=>((G=<I)&J)) & J=>K & (K&L)=>A,
                     new3(K,G,L,H,I,J), new7(M,N,A,G,L,T,P,K,G,B,C).
T5. false :- A & ~B, new3(B,C,D,E,F,A).                        /* --- folded from clause GR */           

/* --------- T6-T8: clauses derived for the snoc contract GS --------- */
T6. new2(A,B,C,D,E,F,G,H,I) :- D=I & I=J & A & B=J & C=(K=>(J>=L & M)) & E & 
         & ~F & G=0 & H & M & ~K & L=0.      
T7. new2(A,B,C,D,E,F,G,H,I) :- D=I & D=J & I=K & K=J & L=M & A & B=M & C=(N=>(M>=V & P)) & 
         & E=(D=<L & Q) & F & G=L & H=(R=>(L>=S & T)) & (T & Q)=>P, new2(N,V,P,J,Q,R,S,T,K).     
T8. false :- A=B & ~((C & D)=>E), new2(F,G,E,B,D,H,I,C,A).     /* --- folded from clause GS */  
\end{verbatim}
\end{sizepar}
\vspace{-1.5mm}
\caption{The set \ttsi{TransfReverseCHCs} of clauses obtained 
by Algorithm~\Cata~\cite{DeAngelisFPP22} 
at the end of Step~(ii).
\label{fig:TransfReverseCHCs-noADTs}}
\end{figure}

It is not really important that the programmer understands the meaning of the newly introduced
predicates, as they are obtained in a fully automatic way by an algorithm whose soundness
is guaranteed in all cases.
It is only important to note that all variables 
in \ttsi{TransfReverseCHCs} are of sort boolean or integer. Indeed, Algorithm~\Cata\ 
always terminates and 
generates a set of clauses without ADT variables in the case where, as in our 
set {\ttsi{ReverseCHCs}} of clauses, the contracts 
are specified by means of {\it catamorphisms}~\cite{DeAngelisFPP22}, that is, 
total functions defined by a simple recursion schema on the ADT structure.
(The notion of catamorphism used here is an adaptation to CHCs of the one popularized
by Meijer et al.~\cite{MeijerFP91} in the area of functional programming.)

The kind of catamorphisms we have used in our example are all instances of the list
catamorphism schema \tts{h} for CHCs depicted in Figure~\ref{fig:listcatamorphisms}. 
The recursion of predicate \tts{h} is on its second argument, which has sort list. 
We leave to the reader to check that, indeed, the predicates
\tts{is\_asorted}, \tts{is\_dsorted}, \tts{hd}, and \tts{leq\_all} (see 
Figure~\ref{fig:RevCHCs}) that we have used for specifying the contracts of \tts{rev} and \tts{snoc}, are all list catamorphisms. In particular, \tts{leq\_all} is a list catamorphism according to the schema of Figure~\ref{fig:listcatamorphisms} by taking \tts{f(X,T,Rf)} to be the atom \tts{true}, and \tts{c(N,X,Rf,B1,B)}
to be defined by the constraint \mbox{`\tts{B = (N=<X \& B1)}'.}


\begin{figure}[ht!]
\begin{sizepar}{10}{12}
\begin{verbatim}
   h(X,[],Res) :- Res=b.           
   h(X,[H|T],Res) :- f(X,T,Rf), h(X,T,R), c(X,H,Rf,R,Res).
\end{verbatim}
\end{sizepar}
\vspace{-3mm}
\caption{Clauses defining the list catamorphism schema \tts{h}.
\label{fig:listcatamorphisms}}
\end{figure}

 \noindent
In Figure~\ref{fig:listcatamorphisms}, we assume that: (i)~\tts{f} is a catamorphism defined
by an instance of the same schema of that figure, (ii)~the second arguments of \tts{h} and \tts{f}
are of sort list, while all other arguments are of basic sort (either boolean or integer), and (iii)~the predicate \tts{c} 
defines a total function
from its first four arguments to its last one.

As already mentioned, we have that if clauses \tts{T1}--\tts{T5} 
are satisfiable then the contract for \tts{rev} is valid and, likewise,
if clauses \tts{T6}--\tts{T8} are satisfiable then the contract for \tts{snoc} 
is valid.

\smallskip
Having derived the set \ttsi{TransfReverseCHCs} of clauses without ADT 
variables, we are ready to perform Step~(iii) of our technique. 
Thus, we invoke a CHC solver (in our case 
SPACER) which, hopefully, is capable to prove the satisfiablity
of \ttsi{TransfReverseCHCs}, because in this set of clauses there are 
variables of basic sort only. 
Indeed, SPACER given clauses \tts{T1}--\tts{T8}, returns the answer `\tts{sat}' stating that 
\mbox{\ttsi{TransfReverseCHCs}} is satisfiable.
At this point, having proved the satifiability of \ttsi{TransfReverseCHCs}, we have proved the validity of 
the contracts for \tts{rev} and \tts{snoc}.


Now we know that the contracts for \tts{rev} and \tts{snoc}
are valid, even if the contract for \tts{rev} is not provable by 
{\sc Stainless}.
However, it is desirable to perform a further step (it is Step~(iv) of our technique that we will describe in the next section) and
derive strengthened contracts for \tts{rev} and \tts{snoc} whose
validity can be automatically shown by \mbox{\sc Stainless}, without appealing 
to an external verification system based on the first three steps of our technique. 
The need for those strengthened contracts comes from the
software engineering
requirement of having a single framework (Scala, in our case) 
{where one specifies programs and contracts, and also proves contract validity
(using {\sc Stainless}, in our case).}
{In the next section we will show how to comply with this
requirement.

Note also that, with respect to the given contracts, the strengthened contracts provide a more 
detailed documentation of the program at hand, and they  
allow the programmer to better understand the 
correctness of the program, as the transformation-based proofs of the 
contracts are often hard to follow.}



\section{Strengthening Program Contracts using CHC Models}
\label{sec:StrengthContracts}
Now we show how to derive strengthened
contracts for \tts{rev} and \tts{snoc} so that the {\sc Stainless} 
verifier can prove their validity.
This derivation can be done by:

\hangindent=5mm
\noindent
(1)~taking into account the definitions of the new predicates introduced by \vericat
during the CHC transformation of Step~(ii), and 
 
\noindent\hangindent=5mm
(2)~constructing the models
of those new predicates as an outcome of the satisfiability proof of the CHCs 
performed by {\sc{SPACER}} at Step~(iii).



\smallskip
\noindent
As we have mentioned at the end of Section~\ref{sec:VerifContracts}, since {\sc Stainless}
is unable to show the contract for \tts{rev}, while it is able to show the contract for \tts{snoc},
we proceed by explaining how to get strengthened contracts by considering only the models of
clauses \tts{T1}--\tts{T5} relative to the \tts{rev} contract. Those clauses refer to the new  
predicates \tts{new3} and \tts{new7} which \vericat introduced
during Step~(ii) by the following definition clauses (modulo variable renaming):



\begin{sizepar}{10}{11}
\begin{verbatim}
D3. new3(BR,N,B,IsDef,Hd,BL) :- is_asorted(L,BL), leq_all(N,Res,B), hd(L,IsDef,Hd), 
               rev(L,Res), is_dsorted(Res,BR).   
               
D7. new7(A1,B1,BC,X,BE,F1,G1,BA,X,J1,K1) :- hd(L,F1,G1), hd(Res,A1,B1), 
               is_dsorted(L,BA), leq_all(X,L,BE), snoc(L,X,Res), is_dsorted(Res,BC), 
               leq_all(J1,Res,K1).
\end{verbatim}   
\end{sizepar}

\noindent
(Note that, in order to eliminate the ADT variables, in the above clauses \tts{D3} and 
\tts{D7}, the arguments of their head atoms are the variables of basic sort occurring in their associated bodies.)

At Step~(iii) {\sc{SPACER}} shows the satisfiability of clauses \tts{T1}--\tts{T5}
by constructing models for \tts{new3} and \tts{new7}.
The model for \tts{new3(BR,N,B,IsDef,Hd,BL)} is: \nopagebreak

\smallskip

\noindent
\tts{M3. (\textasciitilde IsDef => (BR \& B))  ~\&~  (BL => (BR \& ((Hd>=N) => B)))} \hfill (model for \tts{new3}) \hspace*{8mm}

\smallskip

\noindent 
and the model for \tts{new7(A1,B1,BC,X,BE,F1,G1,BA,X,J1,K1)} is: \nopagebreak

\smallskip

\noindent
\tts{M7. (BE \& (X >= J1)) => K1}. \hfill (model for \tts{new7}) \hspace*{8mm}

\smallskip\noindent
First, we consider the definition clause \tts{D3} and the model \tts{M3} for \tts{new3}.
From the left conjunct  \tts{(\textasciitilde IsDef => (BR \& B))} of the model 
we have that if the list \tts{l} is empty, being \tts{IsDef=false}  
(see the clauses for \tts{hd} in Figure~\ref{fig:RevCHCs}), then: (i)~the 
reversed list \tts{res} is empty, (ii)~\tts{res} is vacuously sorted in descending order, 
and (iii)~\tts{leq_all(n,res)} vacuously holds
for all integers \tts{n}. Hence, we get the following  formula {(using the Scala syntax)}, where
the variable \tts{n} has been universally quantified (indeed, it is 
neither an input nor an output variable of \tts{rev}):

\vspace{1mm}
\noindent
\tts{(3.1)~~~forall((n:\,BigInt) => (!(hd(l)._1) ==> (is_dsorted(res) \&\& leq_all(n,res))))}

\vspace{1mm}
\noindent
From the second conjunct \tts{(BL => (BR \& ((Hd>=N) => B)))} of the model 
\tts{M3} for \tts{new3}, we get the following formula (again the variable \tts{n} has been universally quantified):

\vspace{-.4mm}
\begin{sizepar}{10}{11}
\begin{verbatim}
(3.2)   forall((n: BigInt) => (is_asorted(l) ==> 
                         (is_dsorted(res) && ((hd(l)._2 >= n) ==> leq_all(n,res)))))
\end{verbatim}   
\end{sizepar}

\vspace{-1mm}
\noindent
Now the conjunction of the given postconditon of the function \tts{rev}, that is, \tts{is_dsorted(res)} {(see the 
definition of \tts{rev} in Figure~\ref{fig:RevScala})}, and formulas 
\tts{(3.1)} and \tts{(3.2)} 
can be suitably simplified by taking into account the precondition of \tts{rev}, that is, \tts{is_asorted(l)}. After that simplification, we eventually get the strengthened postcondition for \tts{rev} shown in Figure~\ref{fig:strengthpost-rev-1}.


\begin{figure}[ht!]
\vspace{1mm}
\begin{sizepar}{10}{11}
\begin{verbatim}                                
... ensuring { res => is_dsorted(res) &&                            
               forall((n: BigInt) => ((!(hd(l)._1) ==> leq_all(n,res))  &&                        
                                       ((hd(l)._2 >= n) ==> leq_all(n,res)))) }    
\end{verbatim}   
\end{sizepar}
\vspace{-3mm}
\caption{Strengthened postcondition for \tts{rev}. Version 1.
\label{fig:strengthpost-rev-1}}
\end{figure}

\noindent
Then, it remains to consider the definition clause \tts{D7} and the model \tts{M7}  for 
\tts{new7}. 
Now \tts{BE} can be replaced by \tts{true}, because \tts{BE} occurs in the atom 
\mbox{\tts{leq_all(X,L,BE)}} in the body of clause \tts{D7} and by the precondition of \tts{snoc}
(see Figure~\ref{fig:RevScala}), we have that \tts{leq_all(x,l)} holds.  
Moreover, in the two atoms \tts{snoc(L,X,Res)} and \tts{leq_all(J1,Res,K1)} 
 occurring in the body of clause \tts{D7}, the value of the variable~\tts{Res} is that of 
\tts{res} of the Scala function 
\tts{snoc}, and the value of the variable~\tts{K1} is the result of the 
Scala function \mbox{\tts{leq_all(j1,res)}}. Hence, from the model for
\tts{new7}, we get the following formula to be used for 
strengthening the postcondition for \tts{snoc}:\nopagebreak

\vspace{1mm}
\noindent
\tts{(7)~~~forall((j1:\ BigInt) => ((x >= j1)) ==> leq_all(j1,res)))}

\vspace{1mm}
\noindent
where the variable \tts{j1} has been universally quantified, because it is neither an 
input nor an output variable of \tts{snoc}. Thus, the conjunction of the given postcondition of 
\tts{snoc}, that is, \tts{is_dsorted(res)} {(see the 
definition of \tts{snoc} in Figure~\ref{fig:RevScala})}, and formula 
\tts{(7)}, gives us the strengthened postcondition for \tts{snoc} shown in Figure~\ref{fig:strengthpost-snoc}.

\begin{figure}[ht!]
\vspace{1mm}
\begin{sizepar}{10}{11}
\begin{verbatim}                                 
... ensuring { res => is_dsorted(res) &&            
               forall((j1: BigInt) => ((x >= j1) ==> leq_all(j1,res))) } 
\end{verbatim}   
\end{sizepar}
\vspace{-2.5mm}
\caption{Strengthened postcondition for \tts{snoc}.
\label{fig:strengthpost-snoc}}
\end{figure}

By using these new, strengthened postconditions of the contracts for \tts{rev} and 
\tts{snoc}, instead of the
old postconditions, we have that {\sc Stainless} is able to prove
the correctness of both contracts, as desired. 

\smallskip

Let us briefly discuss the soundness of our technique for strengthening
contracts.
We consider the class of contracts specified by catamorphisms 
considered in previous work~\cite{DeAngelisFPP22},
where the termination of the transformation algorithm~\Cata\ is guaranteed.
Soundness of our technique will be shown by proving that, if a given contract is valid and
a CHC solver is able to prove the satisfiability of
the set of clauses obtained by~\Cata, 
then also the strengthened version of the contract, which is derived by using our technique, is valid.

For reasons of simplicity, let us assume that we want to 
show the validity of a contract for 
the function $\mathtt{f}: \alpha \rightarrow \beta$, where $\alpha$ and $\beta$ are ADTs.
Since the contract is specified by catamorphisms,
its validity can be expressed by a formula of the form:

\smallskip

$\mathtt{\forall\, x,y,v_1,v_2.\ (f(x)\! = \!y \,\wedge\, 
p_1(x) \!= \! v_1 \,\wedge\, p_2(y)\! =\! v_2 ~\rightarrow~ c(v_1,v_2))}$ \hfill({\tt C})\hspace{1cm}

\smallskip
\noindent
where $\mathtt{p_1}$ and $\mathtt{p_2}$ are (tuples of) 
catamorphisms and $\mathtt{c(v_1,v_2)}$ is a constraint
on integer and boolean variables. For instance, in our 
\ttsi{Reverse} example, {\tt C} is the formula  

\smallskip
$\mathtt{\forall\, l,res,b1,b2.\ {\tt (rev(l) \!=\! res} \,\wedge\, is\_asorted(l) \!=\! b1 \,\wedge\, is\_dsorted(res) \!=\! b2 ~\rightarrow~ (b1 \rightarrow b2))}$

\smallskip
\noindent
Let us assume that the contract is valid. Thus, if we consider: 
(i)~the set~$P$ of CHCs that translates the program including the function~{\tt f},
and (ii)~the following goal~{\tt{G}} that translates the 
contract (where, as in Figure~\ref{fig:RevCHCs}, 
 `{\tt \textasciitilde}' 
denotes boolean negation):

\smallskip

${\tt G}$: \ $\mathtt{false} \ \tts{:-}\ ${\tt \textasciitilde}$\mathtt{c(V1,V2),\ f(X,Y),\ p_1(X,V1),\ p_2(Y,V2).}$

\smallskip
\noindent
then the set $P \cup \{{\tt G}\}$ of CHCs is satisfiable. 

Algorithm \Cata\  introduces a new definition of the form:

\smallskip

{\tt D}: \ $\mathtt{newf(V1,W1,V2,W2)\  \tts{:-}\ f(X,Y),
\ p_1(X,V1),\ q_1(X,W1),\ p_2(Y,V2),\ q_2(Y,W2).}$

\smallskip

\noindent
by using the catamorphisms $\mathtt{p_1}$ and $\mathtt{p_2}$ in ${\tt G}$ and adding (zero or more) new catamorphisms such as~$\mathtt{q_1}$ 
and~$\mathtt{q_2}$
coming from the contracts for other functions occurring in the program at hand. 
Let $P' \cup \{{\tt G'}\}$ be set of CHCs produced as 
output by algorithm \Cata.  
Some of the CHCs in $\mathtt{P'}$  
have $\mathtt{newf}$ as the head predicate and $\mathtt{G'}$ 
is the goal derived from goal~{\tt G}.
%
%
%
%

If $P' \cup \{{\tt G'}\}$ is proved to be satisfiable by any CHC solver,
then there exists a model of $P' \cup \{{\tt G'}\}$
where the interpretation of $\mathtt{newf(V1,W1,V2,W2)}$ can be expressed as a 
constraint $\mathtt{d(V1,W1,V2,W2)}$, and hence

\smallskip

$M(P') \models \mathtt{\forall\ V1,W1,V2,W2.~~ newf(V1,W1,V2,W2) ~\rightarrow~ d(V1,W1,V2,W2)}$

\smallskip

\noindent
where, for any given set $S$ of definite clauses (i.e., clauses whose head is different from {\tt false}), $M(S)$ denotes the least model of $S$. 
Now, by the soundness of the transformation~\cite{DeAngelisFPP22},

%

\smallskip

$M(P\cup \{{\tt D}\}) \models \mathtt{\forall\ V1,W1,V2,W2.~~  newf(V1,W1,V2,W2) ~\rightarrow~ d(V1,W1,V2,W2)}$

\smallskip

\noindent
and, by using clause {\tt D}, we get:

\smallskip

$M(P) \models \mathtt{\forall\ X,Y,V1,W1,V2,W2.\  
f(X,Y),\, p_1(X,V1),\, q_1(X,W1),\, p_2(Y,V2),\, q_2(Y,W2) ~\rightarrow~ d(V1,W1,V2,W2)}$

\smallskip

\noindent
By using also goal ${\tt G}$, we get:

\smallskip

$M(P) \models \mathtt{\forall\ X,Y,V1,W1,V2,W2.\  
f(X,Y),\ p_1(X,V1),\ q_1(X,W1),\ p_2(Y,V2),\ q_2(Y,W2) ~\rightarrow~}$
\nopagebreak

\hspace{46.5mm}$\mathtt{c(V1,V2) ~\mbox{\tt \&}~ d(V1,W1,V2,W2)}$~~
\hfill({\tt SC})\hspace{1cm}

\smallskip

\noindent
which shows the validity of a strengthened contract for {\tt f}.
Indeed, since catamorphisms are total functions, 
the atoms  $\mathtt{p_1(X,V1)}$, $\mathtt{q_1(X,W1)}$, $\mathtt{p_2(Y,V2)}$, and $\mathtt{q_2(Y,W2)}$ are satisfiable,
for all values of $\mathtt{X}$ and~$\mathtt{Y}$, and the conclusion of {\tt SC} is strengthened with respect to the conclusion of~{\tt C}.
Notice that, however, even if the strengthened contract is valid, we have no guarantee that the verifier (e.g., {\sc Stainless})
is able to prove it.

\smallskip

{Finally, let us make a remark on the \emph{minimality} of the strengthened contracts, related to the fact that}
the information provided by the models of 
the new predicates introduced by algorithm \Cata\ during program transformation, can also be
used, so to say, in a partial way. 
Let us explain this point by 
referring to our program {\ttsi{Reverse}}.
In this case,
instead of the strengthened postcondition for
\tts{rev} that is derived from the conjunction of the given postcondition \tts{is_dsorted(res)}  and 
formulas~\tts{(3.1)} and~\tts{(3.2)}, we could have used the postcondition shown in 
Figure~\ref{fig:strengthpostpost-rev-2}, which is derived from the given postcondition 
and formula~\tts{(3.2)} only. 

\begin{figure}[ht!]
\vspace{1mm}
\begin{sizepar}{10}{11}
\begin{verbatim}                                 
... ensuring { res => is_dsorted(res) &&            
               forall((n: BigInt) => (((hd(l)._2 >= n) ==> leq_all(n,res)))) }
\end{verbatim}   
\end{sizepar}
\vspace{-2mm}
\caption{Strengthened postcondition for \tts{rev}. Version 2.
\label{fig:strengthpostpost-rev-2}}
\vspace{-2mm}
\end{figure}

This postcondition is sufficiently strong to allow the {\sc Stainless} verifier to
prove the contracts for \tts{rev} and \tts{snoc} (actually, {\sc Stainless} does 
take for that proof less
time with respect to the time taken for the more complex postcondition of Figure~\ref{fig:strengthpost-rev-1}). 
However, in general, the 
strengthened postconditions that do not take into account the whole
information which is derivable from the model of new predicates, provide 
a less informative description of the behaviour of the program functions at hand.
We leave for future work the issue of computing {minimally strengthened} postconditions.
%
%
%
%

\section{Conclusions}
\label{sec:Conclusions}
A good software engineering practice requires us 
to associate a contract with every program function, that is, to write, for each program function, a 
precondition and a postcondition.

The current software technology provides, together with compilers (and 
interpreters), also
program verifiers, so that given a program function and its contract, one can execute the 
function and also prove {its partial or total correctness with respect to its contract.}
%
{Due to undecidability results, there is no program verifier that is able to 
prove (or disprove) program correctness in all cases.}
However, often a verifier is not able to show correctness simply because
the contracts have not been specified in an adequate manner.

We have addressed this problem and we have proposed a 
technique 
which is capable of improving the contracts, and in particular, it is capable 
of strengthening their postconditions, so that a given verifier is successful in
proving the correctness of functions, while it is not successful when
trying to prove their correctness using the original postconditons.

We have considered programs and contracts written in the functional fragment of
Scala and we have considered the {\sc Stainless}~\cite{HamzaVK19} verifier
for Scala programs.
Our technique is based on the translation of the program functions 
and their contracts for which  {\sc Stainless} is not successful,
into a set of constrained Horn clauses (CHCs)~\cite{DeAngelisFGHPP21,Gr&12}. Then, in the case where programs
manipulate Algebraic Data Structures, those clauses are 
transformed by \vericat~\cite{DeAngelisFPP22}, so that their satisfiability can hopefully 
be proved by {\sc SPACER} (or a different CHC solver) in the domain of integers and/or booleans. 
If satisfiability is proved and
a model for those clauses is found (which is defined by constraints on integers and/or
booleans), then via a final translation step, 
we derive from that model suitable 
strengthened postconditions for the contracts. 
These derived contracts are guaranteed to be valid and can hopefully be proved by 
the {\sc Stainless} verifier. However, at the moment,
we do not have any general result characterizing these successful cases. 

The derivation of strengthened contracts is an important objective 
from a software
engineering point of view, because strengthened contracts 
provide a uniform framework
(the Scala framework, in our case) where to write programs and their
contracts. In this uniform framework an automatic system, such as 
{\sc Stainless}, can show that the given contracts are valid. This is an
important feature, because contracts can be viewed both as a documentation 
of the programs and also as a specification of their behavior.

As mentioned above, the proof of satisfiability of the CHCs derived 
after translation from the given programs, is already a proof of the contracts, 
but that proof is not given in the same framework. Moreover, it should rely on
the correctness of the translation from Scala programs to CHCs
and also on the correctness of the transformation for ADT removal. 
In addition, the derivation of strengthened postconditions and their proof 
done by {\sc Stainless} increase the reliability
of the validity proofs of the contracts, in the sense that the satisfiability preserving transformation performed by \vericat and the
model constructed by the CHC solver {\sc SPACER} are shown to 
agree with the behaviour 
of the {\sc Stainless} Scala verifier.
This agreement of automatic tools is very important in practice, 
and in particular when we deal 
with large programs.

Currently, we are working towards the full
mechanization of the two steps of our technique that 
are still performed manually, even though they are performed 
in a systematic way, namely: 
(i)~the semantic preserving translation from Scala programs into constrained Horn clauses,
and (ii)~the construction of strengthened postconditons 
from the models of the satisfiable clauses provided by a CHC solver (such as 
{\sc{SPACER}} or Eldarica).

\section*{Acknowledgments}
The authors warmly thank the anonymous reviewers for their helpful comments and suggestions.
The authors are members of the INdAM Research Group GNCS.


\begin{thebibliography}{10}
\providecommand{\bibitemdeclare}[2]{}
\providecommand{\surnamestart}{}
\providecommand{\surnameend}{}
\providecommand{\urlprefix}{Available at }
\providecommand{\url}[1]{\texttt{#1}}
\providecommand{\href}[2]{\texttt{#2}}
\providecommand{\urlalt}[2]{\href{#1}{#2}}
\providecommand{\doi}[1]{doi:\urlalt{http://dx.doi.org/#1}{#1}}
\providecommand{\eprint}[1]{arXiv:\urlalt{https://arxiv.org/abs/#1}{#1}}
\providecommand{\bibinfo}[2]{#2}

\bibitemdeclare{incollection}{Boogie}
\bibitem{Boogie}
\bibinfo{author}{M.~\surnamestart Barnett\surnameend},
  \bibinfo{author}{B.-Y.~E. \surnamestart Chang\surnameend},
  \bibinfo{author}{R.~\surnamestart {De Line}\surnameend},
  \bibinfo{author}{B.~\surnamestart Jacobs\surnameend} \&
  \bibinfo{author}{K.~R.~M. \surnamestart Leino\surnameend}
  (\bibinfo{year}{2006}): \emph{\bibinfo{title}{Boogie: {A} Modular Reusable
  Verifier for Object-Oriented Programs}}.
\newblock In \bibinfo{editor}{F.~\surnamestart de~Boer\surnameend},
  \bibinfo{editor}{M.~M. \surnamestart Bonsangue\surnameend},
  \bibinfo{editor}{S.~\surnamestart Graf\surnameend} \& \bibinfo{editor}{W.-P.
  \surnamestart de~Roever\surnameend}, editors: {\sl \bibinfo{booktitle}{Formal
  Methods for Components and Objects}}, \bibinfo{series}{Lecture Notes in
  Computer Science 4111}, \bibinfo{publisher}{Springer}, pp.
  \bibinfo{pages}{364--387}, \doi{10.1007/11804192_17}.

\bibitemdeclare{inproceedings}{CVC4}
\bibitem{CVC4}
\bibinfo{author}{C.~\surnamestart Barrett\surnameend}, \bibinfo{author}{C.~L.
  \surnamestart Conway\surnameend}, \bibinfo{author}{M.~\surnamestart
  Deters\surnameend}, \bibinfo{author}{L.~\surnamestart Hadarean\surnameend},
  \bibinfo{author}{D.~\surnamestart Jovanovic\surnameend},
  \bibinfo{author}{T.~\surnamestart King\surnameend},
  \bibinfo{author}{A.~\surnamestart Reynolds\surnameend} \&
  \bibinfo{author}{C.~\surnamestart Tinelli\surnameend} (\bibinfo{year}{2011}):
  \emph{\bibinfo{title}{{CVC4}}}.
\newblock In \bibinfo{editor}{Ganesh \surnamestart Gopalakrishnan\surnameend}
  \& \bibinfo{editor}{Shaz \surnamestart Qadeer\surnameend}, editors: {\sl
  \bibinfo{booktitle}{23rd {CAV}~'11}}, \bibinfo{series}{Lecture Notes in
  Computer Science 6806}, \bibinfo{publisher}{Springer}, pp.
  \bibinfo{pages}{171--177}, \doi{10.1007/978-3-642-22110-1_14}.

\bibitemdeclare{book}{BoochB94}
\bibitem{BoochB94}
\bibinfo{author}{Grady \surnamestart Booch\surnameend} \& \bibinfo{author}{Doug
  \surnamestart Bryan\surnameend} (\bibinfo{year}{1994}):
  \emph{\bibinfo{title}{Software engineering with Ada {(3.} ed.)}}.
\newblock \bibinfo{series}{Benjamin/Cummings series in object-oriented software
  engineering}, \bibinfo{publisher}{Benjamin/Cummings}.

\bibitemdeclare{inproceedings}{MaS13}
\bibitem{MaS13}
\bibinfo{author}{Alessandro \surnamestart Cimatti\surnameend},
  \bibinfo{author}{Alberto \surnamestart Griggio\surnameend},
  \bibinfo{author}{Bastiaan \surnamestart Schaafsma\surnameend} \&
  \bibinfo{author}{Roberto \surnamestart Sebastiani\surnameend}
  (\bibinfo{year}{2013}): \emph{\bibinfo{title}{The {MathSAT5 SMT} solver}}.
\newblock In \bibinfo{editor}{Nir \surnamestart Piterman\surnameend} \&
  \bibinfo{editor}{Scott \surnamestart Smolka\surnameend}, editors: {\sl
  \bibinfo{booktitle}{19th TACAS~'13}}, \bibinfo{series}{Lecture Notes in
  Computer Science 7795}, \bibinfo{publisher}{Springer}, pp.
  \bibinfo{pages}{93--107}, \doi{10.1007/978-3-642-36742-7_7}.

\bibitemdeclare{article}{DeAngelisFGHPP21}
\bibitem{DeAngelisFGHPP21}
\bibinfo{author}{E.~\surnamestart {De Angelis}\surnameend},
  \bibinfo{author}{F.~\surnamestart Fioravanti\surnameend},
  \bibinfo{author}{J.~P. \surnamestart Gallagher\surnameend},
  \bibinfo{author}{M.~V. \surnamestart Hermenegildo\surnameend},
  \bibinfo{author}{A.~\surnamestart Pettorossi\surnameend} \&
  \bibinfo{author}{M.~\surnamestart Proietti\surnameend}
  (\bibinfo{year}{2021}): \emph{\bibinfo{title}{Analysis and Transformation of
  Constrained {H}orn Clauses for Program Verification}}.
\newblock {\sl \bibinfo{journal}{Theory and Practice of Logic Programming}},
  pp. \bibinfo{pages}{1--69}, \doi{10.1017/S1471068421000211}.

\bibitemdeclare{article}{De&18a}
\bibitem{De&18a}
\bibinfo{author}{E.~\surnamestart {De Angelis}\surnameend},
  \bibinfo{author}{F.~\surnamestart Fioravanti\surnameend},
  \bibinfo{author}{A.~\surnamestart Pettorossi\surnameend} \&
  \bibinfo{author}{M.~\surnamestart Proietti\surnameend}
  (\bibinfo{year}{2018}): \emph{\bibinfo{title}{Solving {H}orn Clauses on
  Inductive Data Types Without Induction}}.
\newblock {\sl \bibinfo{journal}{Theory and Practice of Logic Programming}}
  \bibinfo{volume}{18}(\bibinfo{number}{3-4}), pp. \bibinfo{pages}{452--469},
  \doi{10.1017/S1471068418000157}.

\bibitemdeclare{inproceedings}{De&20a}
\bibitem{De&20a}
\bibinfo{author}{E.~\surnamestart {De Angelis}\surnameend},
  \bibinfo{author}{F.~\surnamestart Fioravanti\surnameend},
  \bibinfo{author}{A.~\surnamestart Pettorossi\surnameend} \&
  \bibinfo{author}{M.~\surnamestart Proietti\surnameend}
  (\bibinfo{year}{2020}): \emph{\bibinfo{title}{Removing Algebraic Data Types
  from Constrained {H}orn Clauses Using Difference Predicates}}.
\newblock In \bibinfo{editor}{N.~\surnamestart Peltier\surnameend} \&
  \bibinfo{editor}{V.~\surnamestart Sofronie-Stokkermans\surnameend}, editors:
  {\sl \bibinfo{booktitle}{Proceedings of the International Joint Conference on
  Automated Reasoning, IJCAR~2020}}, \bibinfo{series}{Lecture Notes in
  Artificial Intelligence 12166}, \bibinfo{publisher}{Springer}, pp.
  \bibinfo{pages}{83--102}, \doi{10.1007/978-3-030-51074-9_6}.

\bibitemdeclare{article}{De&22a}
\bibitem{De&22a}
\bibinfo{author}{E.~\surnamestart {De Angelis}\surnameend},
  \bibinfo{author}{F.~\surnamestart Fioravanti\surnameend},
  \bibinfo{author}{A.~\surnamestart Pettorossi\surnameend} \&
  \bibinfo{author}{M.~\surnamestart Proietti\surnameend}
  (\bibinfo{year}{2022a}): \emph{\bibinfo{title}{Satisfiability of constrained
  {H}orn clauses on algebraic data types: {A} transformation-based approach}}.
\newblock {\sl \bibinfo{journal}{Journal of Logic and Computation}}
  \bibinfo{volume}{32}, pp. \bibinfo{pages}{402--442},
  \doi{10.1093/logcom/exab090}.

\bibitemdeclare{article}{DeAngelisFPP22}
\bibitem{DeAngelisFPP22}
\bibinfo{author}{E.~\surnamestart {De Angelis}\surnameend},
  \bibinfo{author}{M.~\surnamestart Proietti\surnameend},
  \bibinfo{author}{F.~\surnamestart Fioravanti\surnameend} \&
  \bibinfo{author}{A.~\surnamestart Pettorossi\surnameend}
  (\bibinfo{year}{2022}): \emph{\bibinfo{title}{Verifying Catamorphism-Based
  Contracts using Constrained {H}orn Clauses}}.
\newblock {\sl \bibinfo{journal}{Theory and Practice of Logic Programming}}
  \bibinfo{volume}{22}(\bibinfo{number}{4}), pp. \bibinfo{pages}{555--572},
  \doi{10.1017/S1471068422000175}.

\bibitemdeclare{inproceedings}{Why3}
\bibitem{Why3}
\bibinfo{author}{J.-C. \surnamestart Filli{\^{a}}tre\surnameend} \&
  \bibinfo{author}{A.~\surnamestart Paskevich\surnameend}
  (\bibinfo{year}{2013}): \emph{\bibinfo{title}{Why3 - {W}here Programs Meet
  Provers}}.
\newblock In \bibinfo{editor}{M.~\surnamestart Felleisen\surnameend} \&
  \bibinfo{editor}{Ph. \surnamestart Gardner\surnameend}, editors: {\sl
  \bibinfo{booktitle}{Programming Languages and Systems, 22nd European
  Symposium on Programming, {ESOP}'13, Rome, Italy, March 16--24, 2013}},
  \bibinfo{series}{Lecture Notes in Computer Science 7792},
  \bibinfo{publisher}{Springer}, pp. \bibinfo{pages}{125--128},
  \doi{10.1007/978-3-642-37036-6_8}.

\bibitemdeclare{inproceedings}{Flo67}
\bibitem{Flo67}
\bibinfo{author}{R.~W. \surnamestart Floyd\surnameend} (\bibinfo{year}{1967}):
  \emph{\bibinfo{title}{Assigning Meanings to Programs}}.
\newblock In \bibinfo{editor}{J.~T. \surnamestart Schwartz\surnameend}, editor:
  {\sl \bibinfo{booktitle}{Proceedings of Symposium on Applied Mathematics,
  Vol. 19}}, \bibinfo{publisher}{American Mathematical Society, Providence,
  R.I., USA}, pp. \bibinfo{pages}{19--32}, \doi{10.1007/978-94-011-1793-7_4}.

\bibitemdeclare{article}{GovindSG22}
\bibitem{GovindSG22}
\bibinfo{author}{H.~\surnamestart {Govind V. K.}\surnameend},
  \bibinfo{author}{S.~\surnamestart Shoham\surnameend} \&
  \bibinfo{author}{A.~\surnamestart Gurfinkel\surnameend}
  (\bibinfo{year}{2022}): \emph{\bibinfo{title}{Solving constrained {H}orn
  clauses modulo algebraic data types and recursive functions}}.
\newblock {\sl \bibinfo{journal}{Proc. {ACM} Program. Lang.}}
  \bibinfo{volume}{6}(\bibinfo{number}{{POPL}}), pp. \bibinfo{pages}{1--29},
  \doi{10.1145/3498722}.

\bibitemdeclare{inproceedings}{Gr&12}
\bibitem{Gr&12}
\bibinfo{author}{S.~\surnamestart Grebenshchikov\surnameend},
  \bibinfo{author}{N.~P. \surnamestart Lopes\surnameend},
  \bibinfo{author}{C.~\surnamestart Popeea\surnameend} \&
  \bibinfo{author}{A.~\surnamestart Rybalchenko\surnameend}
  (\bibinfo{year}{2012}): \emph{\bibinfo{title}{Synthesizing software verifiers
  from proof rules}}.
\newblock In: {\sl \bibinfo{booktitle}{33rd ACM SIGPLAN Conf.~Programming
  Language Design and Implementation, PLDI~'12}}, pp.
  \bibinfo{pages}{405--416}, \doi{10.1145/2345156.2254112}.

\bibitemdeclare{article}{HamzaVK19}
\bibitem{HamzaVK19}
\bibinfo{author}{J.~\surnamestart Hamza\surnameend},
  \bibinfo{author}{N.~\surnamestart Voirol\surnameend} \&
  \bibinfo{author}{V.~\surnamestart Kuncak\surnameend} (\bibinfo{year}{2019}):
  \emph{\bibinfo{title}{System {FR:} formalized foundations for the {S}tainless
  verifier}}.
\newblock {\sl \bibinfo{journal}{Proc. {ACM} Program. Lang.}}
  \bibinfo{volume}{3}(\bibinfo{number}{{OOPSLA}}), pp.
  \bibinfo{pages}{166:1--166:30}, \doi{10.1145/3360592}.

\bibitemdeclare{article}{Hermenegildo&12}
\bibitem{Hermenegildo&12}
\bibinfo{author}{M.~\surnamestart Hermenegildo\surnameend},
  \bibinfo{author}{F.~\surnamestart Bueno\surnameend},
  \bibinfo{author}{M.~\surnamestart Carro\surnameend},
  \bibinfo{author}{P.~\surnamestart L{\'o}pez-Garc{\'i}a\surnameend},
  \bibinfo{author}{E.~\surnamestart Mera\surnameend}, \bibinfo{author}{J.~F.
  \surnamestart Morales\surnameend} \& \bibinfo{author}{G.~\surnamestart
  Puebla\surnameend} (\bibinfo{year}{2012}): \emph{\bibinfo{title}{{A}n
  Overview of {C}iao and its Design Philosophy}}.
\newblock {\sl \bibinfo{journal}{Theory and Practice of Logic Programming}}
  \bibinfo{volume}{12}(\bibinfo{number}{1--2}), pp. \bibinfo{pages}{219--252},
  \doi{10.1017/S1471068411000457}.

\bibitemdeclare{article}{Hoa69}
\bibitem{Hoa69}
\bibinfo{author}{C.A.R. \surnamestart Hoare\surnameend} (\bibinfo{year}{1969}):
  \emph{\bibinfo{title}{An {A}xiomatic {B}asis for {C}omputer {P}rogramming}}.
\newblock {\sl \bibinfo{journal}{CACM}}
  \bibinfo{volume}{12}(\bibinfo{number}{10}), pp. \bibinfo{pages}{576--580,
  583}, \doi{10.1145/363235.363259}.

\bibitemdeclare{inproceedings}{HoR18}
\bibitem{HoR18}
\bibinfo{author}{H.~\surnamestart Hojjat\surnameend} \& \bibinfo{author}{Ph.
  \surnamestart R{\"{u}}mmer\surnameend} (\bibinfo{year}{2018}):
  \emph{\bibinfo{title}{The {ELDARICA} {H}orn Solver}}.
\newblock In \bibinfo{editor}{N.~\surnamestart Bj{\o}rner\surnameend} \&
  \bibinfo{editor}{A.~\surnamestart Gurfinkel\surnameend}, editors: {\sl
  \bibinfo{booktitle}{Formal Methods in Computer Aided Design,
  \mbox{FMCAD~2018}}}, \bibinfo{publisher}{{IEEE}}, pp. \bibinfo{pages}{1--7},
  \doi{10.23919/FMCAD.2018.8603013}.

\bibitemdeclare{inproceedings}{Ko&14}
\bibitem{Ko&14}
\bibinfo{author}{A.~\surnamestart Komuravelli\surnameend},
  \bibinfo{author}{A.~\surnamestart Gurfinkel\surnameend} \&
  \bibinfo{author}{S.~\surnamestart Chaki\surnameend} (\bibinfo{year}{2014}):
  \emph{\bibinfo{title}{{SMT}-Based Model Checking for Recursive Programs}}.
\newblock In: {\sl \bibinfo{booktitle}{26th {CAV}~'14}},
  \bibinfo{series}{Lecture Notes in Computer Science 8559},
  \bibinfo{publisher}{Springer}, pp. \bibinfo{pages}{17--34},
  \doi{10.1007/978-3-319-08867-9_2}.

\bibitemdeclare{inproceedings}{KostyukovMF21}
\bibitem{KostyukovMF21}
\bibinfo{author}{Yurii \surnamestart Kostyukov\surnameend},
  \bibinfo{author}{Dmitry \surnamestart Mordvinov\surnameend} \&
  \bibinfo{author}{Grigory \surnamestart Fedyukovich\surnameend}
  (\bibinfo{year}{2021}): \emph{\bibinfo{title}{Beyond the elementary
  representations of program invariants over algebraic data types}}.
\newblock In \bibinfo{editor}{Stephen~N. \surnamestart Freund\surnameend} \&
  \bibinfo{editor}{Eran \surnamestart Yahav\surnameend}, editors: {\sl
  \bibinfo{booktitle}{{PLDI} '21: 42nd {ACM} {SIGPLAN} International Conference
  on Programming Language Design and Implementation, Virtual Event, Canada,
  June 20-25, 2021}}, \bibinfo{publisher}{{ACM}}, pp.
  \bibinfo{pages}{451--465}, \doi{10.1145/3453483.3454055}.

\bibitemdeclare{inproceedings}{Lei13}
\bibitem{Lei13}
\bibinfo{author}{K.~R.~M. \surnamestart Leino\surnameend}
  (\bibinfo{year}{2013}): \emph{\bibinfo{title}{Developing Verified Programs
  with {D}afny}}.
\newblock In: {\sl \bibinfo{booktitle}{Intl. Conf. on Software
  Engineering~'13}}, \bibinfo{publisher}{IEEE Press}, pp.
  \bibinfo{pages}{1488--1490}, \doi{10.1109/ICSE.2013.6606754}.

\bibitemdeclare{inproceedings}{MeijerFP91}
\bibitem{MeijerFP91}
\bibinfo{author}{E.~\surnamestart Meijer\surnameend}, \bibinfo{author}{M.~M.
  \surnamestart Fokkinga\surnameend} \& \bibinfo{author}{R.~\surnamestart
  Paterson\surnameend} (\bibinfo{year}{1991}): \emph{\bibinfo{title}{Functional
  Programming with Bananas, Lenses, Envelopes and Barbed Wire}}.
\newblock In \bibinfo{editor}{J.~\surnamestart Hughes\surnameend}, editor: {\sl
  \bibinfo{booktitle}{Functional Programming Languages and Computer
  Architecture, 5th {ACM} Conference, Cambridge, MA, USA, August 26-30, 1991}},
  \bibinfo{series}{Lecture Notes in Computer Science 523},
  \bibinfo{publisher}{Springer}, pp. \bibinfo{pages}{124--144},
  \doi{10.1007/3540543961\_7}.

\bibitemdeclare{book}{Meyer91}
\bibitem{Meyer91}
\bibinfo{author}{Bertrand \surnamestart Meyer\surnameend}
  (\bibinfo{year}{1991}): \emph{\bibinfo{title}{Eiffel: The Language}}.
\newblock \bibinfo{publisher}{Prentice-Hall}.

\bibitemdeclare{inproceedings}{DeB08}
\bibitem{DeB08}
\bibinfo{author}{L.~M. \surnamestart de~Moura\surnameend} \&
  \bibinfo{author}{N.~\surnamestart Bj{\o}rner\surnameend}
  (\bibinfo{year}{2008}): \emph{\bibinfo{title}{Z3: {A}n Efficient {SMT}
  Solver}}.
\newblock In: {\sl \bibinfo{booktitle}{14th {TACAS}~'08}},
  \bibinfo{series}{Lecture Notes in Computer Science 4963},
  \bibinfo{publisher}{Springer}, pp. \bibinfo{pages}{337--340},
  \doi{10.1007/978-3-540-78800-3\_24}.

\bibitemdeclare{book}{OderskySV11}
\bibitem{OderskySV11}
\bibinfo{author}{M.~\surnamestart Odersky\surnameend},
  \bibinfo{author}{L.~\surnamestart Spoon\surnameend} \&
  \bibinfo{author}{B.~\surnamestart Venners\surnameend} (\bibinfo{year}{2011}):
  \emph{\bibinfo{title}{Programming in {S}cala: {A} Comprehensive Step-by-Step
  Guide}}, \bibinfo{edition}{2nd} edition.
\newblock \bibinfo{publisher}{Artima Incorporation},
  \bibinfo{address}{Sunnyvale, CA, USA}.

\bibitemdeclare{inproceedings}{ReK15}
\bibitem{ReK15}
\bibinfo{author}{A.~\surnamestart Reynolds\surnameend} \&
  \bibinfo{author}{V.~\surnamestart Kuncak\surnameend} (\bibinfo{year}{2015}):
  \emph{\bibinfo{title}{Induction for {SMT} Solvers}}.
\newblock In \bibinfo{editor}{Deepak \surnamestart D'Souza\surnameend},
  \bibinfo{editor}{Akash \surnamestart Lal\surnameend} \&
  \bibinfo{editor}{Kim~Guldstrand \surnamestart Larsen\surnameend}, editors:
  {\sl \bibinfo{booktitle}{16th {VMCAI}}}, \bibinfo{series}{Lecture Notes in
  Computer Science 8931}, \bibinfo{publisher}{Springer}, pp.
  \bibinfo{pages}{80--98}, \doi{10.1007/978-3-662-46081-8_5}.

\bibitemdeclare{misc}{Solidity}
\bibitem{Solidity}
\bibinfo{author}{\surnamestart Solidity\surnameend} (\bibinfo{year}{2022}):
  \emph{\bibinfo{title}{Solidity v0.8.12 Documentation}}.
\newblock \bibinfo{howpublished}{\url{https://docs.soliditylang.org/}}.

\bibitemdeclare{incollection}{Su&11}
\bibitem{Su&11}
\bibinfo{author}{{Philippe} \surnamestart Suter\surnameend},
  \bibinfo{author}{A.~S. \surnamestart K\"{o}ksal\surnameend} \&
  \bibinfo{author}{V.~\surnamestart Kuncak\surnameend} (\bibinfo{year}{2011}):
  \emph{\bibinfo{title}{Satisfiability Modulo Recursive Programs}}.
\newblock In \bibinfo{editor}{E.~\surnamestart Yahav\surnameend}, editor: {\sl
  \bibinfo{booktitle}{18th {SAS}~'11}}, \bibinfo{series}{Lecture Notes in
  Computer Science 6887}, \bibinfo{publisher}{Springer}, pp.
  \bibinfo{pages}{298--315}, \doi{10.1007/978-3-642-23702-7\_23}.

\bibitemdeclare{inproceedings}{Un&17}
\bibitem{Un&17}
\bibinfo{author}{H.~\surnamestart Unno\surnameend},
  \bibinfo{author}{S.~\surnamestart Torii\surnameend} \&
  \bibinfo{author}{H.~\surnamestart Sakamoto\surnameend}
  (\bibinfo{year}{2017}): \emph{\bibinfo{title}{Automating Induction for
  Solving {H}orn Clauses}}.
\newblock In \bibinfo{editor}{Rupak \surnamestart Majumdar\surnameend} \&
  \bibinfo{editor}{Viktor \surnamestart Kuncak\surnameend}, editors: {\sl
  \bibinfo{booktitle}{29th {CAV}~'17, Part {II}}}, \bibinfo{series}{Lecture
  Notes in Computer Science 10427}, \bibinfo{publisher}{Springer}, pp.
  \bibinfo{pages}{571--591}, \doi{10.1007/978-3-319-63390-9_30}.

\bibitemdeclare{inproceedings}{Ya&19}
\bibitem{Ya&19}
\bibinfo{author}{Weikun \surnamestart Yang\surnameend},
  \bibinfo{author}{Grigory \surnamestart Fedyukovich\surnameend} \&
  \bibinfo{author}{Aarti \surnamestart Gupta\surnameend}
  (\bibinfo{year}{2019}): \emph{\bibinfo{title}{Lemma Synthesis for Automating
  Induction over Algebraic Data Types}}.
\newblock In \bibinfo{editor}{Thomas \surnamestart Schiex\surnameend} \&
  \bibinfo{editor}{Simon \surnamestart de~Givry\surnameend}, editors: {\sl
  \bibinfo{booktitle}{25th Int. Conf. Principles and Practice of Constraint
  Programming, {CP~2019}}}, \bibinfo{series}{Lecture Notes in Computer Science
  11802}, \bibinfo{publisher}{Springer}, pp. \bibinfo{pages}{600--617},
  \doi{10.1007/978-3-030-30048-7\_35}.

\end{thebibliography}

\end{document}